\documentclass[aps,prl,reprint,showkeys,superscriptaddress,amsmath,amssymb]{revtex4-1}
\usepackage{upgreek}
\usepackage[dvipdfmx]{graphicx} 

\begin{document}
\title{Broadband and energy-concentrating terahertz coherent perfect absorber based on a self-complementary metasurface}

\author{Yoshiro Urade}
\email{urade@giga.kuee.kyoto-u.ac.jp}
\affiliation{Department of Electronic Science and Engineering, Kyoto University, Kyoto 615-8510, Japan}
\author{Yosuke Nakata}
\affiliation{Center for Energy and Environmental Science, Shinshu University, 4-17-1 Wakasato, Nagano 380-8553, Japan}
\author{Toshihiro Nakanishi}
\affiliation{Department of Electronic Science and Engineering, Kyoto University, Kyoto 615-8510, Japan}
\author{Masao Kitano}
\affiliation{Department of Electronic Science and Engineering, Kyoto University, Kyoto 615-8510, Japan}

\date{\today}

\begin{abstract}
We demonstrate that a self-complementary checkerboard-like metasurface works as a broadband coherent perfect absorber~(CPA) when symmetrically illuminated by
 two counter-propagating incident waves.
 A theoretical analysis based on wave interference and results of numerical simulations of the proposed metasurface are provided.
 In addition, we experimentally demonstrate the proposed CPA in the terahertz regime by using a time-domain spectroscopy technique.
 We observe that the metasurface can work as a CPA below its lowest diffraction frequency.
 The size of the absorptive areas of the proposed CPA can be much smaller than the incident wavelength.
 Unlike conventional CPAs, the presented one simultaneously achieves the broadband operation and energy concentration of electromagnetic waves at the deep-subwavelength scale.
\vskip0.5pc \noindent\copyright \, 2016 \hskip.05in Optical Society of America. One print or electronic copy may be made
for personal use only. Systematic reproduction and distribution, duplication of any
material in this paper for a fee or for commercial purposes, or modifications of the
content of this paper are prohibited.
\end{abstract}

\maketitle

Because of destructive interference in all scattering paths, some materials that usually show imperfect absorption can work as perfect electromagnetic absorbers when coherently illuminated by multiple incident waves.
This coherent absorption-enhanced system is known as a coherent perfect absorber~(CPA).
Following their theoretical proposal as a time-reversed process of lasing at threshold~\cite{Chong2010} and experimental demonstrations using a silicon slab cavity~\cite{Wan2011}, CPAs have attracted growing attention from a number of researchers.
It has been theoretically shown that a similar phenomenon occurs for thin absorptive films~(thin-film CPAs), which are extremely thin compared to incident wavelengths and have a wide bandwidth~\cite{Pu2012}. Thin-film CPAs have been experimentally demonstrated in the microwave region~\cite{Li2015c}.
Moreover, multiple systems have been shown to function as CPAs, including monolayer graphene~\cite{Fan2014}, nanostructured graphene films~\cite{Zhang2012}, metasurfaces~(planar metamaterials)~\cite{Kang2013,Zhu2016}, plasmonic waveguides~\cite{Park2015}, and cavity optomechanical systems~\cite{Yan2014}.
CPAs have been considered in the single-photon regime~\cite{Huang2014,Roger2015}.
Using the phase sensitivity of the absorption enhancement in CPAs, all-optical modulators based on CPAs have been proposed~\cite{Zhang2012c,Mock2012}.
Inspired by CPAs, coherent perfect polarization rotation has also been  discussed and experimentally demonstrated~\cite{Crescimanno2012,Zhou2016}.

In our previous study, we theoretically demonstrated that a self-complementary metasurface with $m$-fold rotational symmetry~($m\geq 3$), such as a metallic checkerboard loaded with resistive sheets, functions as a CPA below its lowest diffraction frequency~\cite{Nakata2013}. Here we define a self-complementary metasurface as a metasurface whose structure can be brought into congruence with its complementary structure obtained by the impedance inversion~\cite{BaumNote1974}. The impedance inversion is a natural extension of the metal--hole interchange in Babinet's principle~\cite{JacksonBook} to finite sheet impedance and is defined by $Z^\text{(comp)}_\mathrm{s}(x,y)=Z_0^2/(4n^2Z_\mathrm{s}(x,y))$, where $(x,y)$ denotes the coordinates on the metasurface; $Z_\mathrm{s}(x,y)$ and $Z^\text{(comp)}_\mathrm{s}(x,y)$ are the spatially varying sheet impedance of the metasurface and its complement, respectively; $Z_0\sim 377\,\Omega$ is the impedance of a vacuum; and $n$ is the refractive index of the surrounding dielectric environment~\cite{Nakata2013}.
Scattering problems for the complementary structure obtained via the impedance inversion are related to those for the original structure via Babinet's principle. 

In this Letter, we  theoretically and experimentally investigate a novel CPA using a self-complementary checkerboard-like metasurface~\cite{Nakata2013,Urade2015} in the terahertz frequency region. The proposed CPA is able to not only operate in a broad bandwidth as a thin-film CPA~\cite{Pu2012,Li2015c}, but also to concentrate the energy of incident electromagnetic waves on deep-subwavelength absorptive regions.
The three-dimensional energy concentration realized by the proposed CPA is distinct from thin-film CPAs using unstructured films, whose absorptive regions are subwavelength-sized only in the thickness direction.
Although similar three-dimensional energy concentration can be achieved by exploiting other metamaterial structures~\cite{Tao2008,Zhang2015c} for single-sided excitation, the presented device is distinguished from them by phase sensitivity and by the fact that, in principle, its absorptive regions can be made arbitrarily small without changing the size of the unit cell. 

\begin{figure}[tb]
\centering
\includegraphics[width=7.5cm]{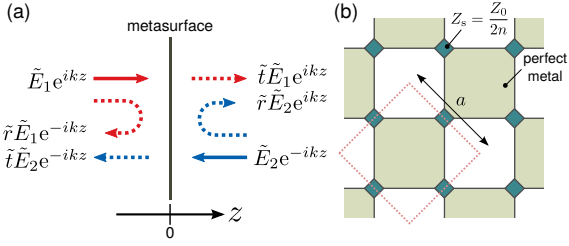}
\caption{(a)~Schematic of the theoretical model. Two counter-propagating waves illuminate a metasurface at $z=0$ in a dielectric medium with refractive index $n$, where $k$ denotes the wavenumber. (b)~Schematic of a self-complementary checkerboard-like metasurface.}
\label{fig:theory}
\end{figure}

First, we review the necessary conditions to realize a CPA by using a simple theoretical model, as shown in Fig.~\ref{fig:theory}(a).
In a dielectric medium with refractive index $n$, there is a single-layer metasurface placed at $z=0$. Two counter-propagating plane waves in the same linear polarization state are normally incident on the metasurface with complex amplitudes of electric fields $\tilde{E}_1$ and $\tilde{E}_2$, assuming a harmonic time dependence $\exp (-i \omega t)$ with an angular frequency $\omega$.
In addition, we assume that the metasurface has periodicity and $m$-fold symmetry ($m\geq 3$), i.e., cross-polarization conversion does not occur~\cite{Nakata2013}.
Furthermore, we consider a frequency range $f<f_0 :=c_0/(na)$, where scattering into higher-order diffraction modes by the metasurface can be ignored. Here, $c_0$ is the speed of light in a vacuum and $a$ is the periodicity of the metasurface.
Let $\tilde{t}$ and $\tilde{r}$ be the zeroth order complex amplitude transmission and reflection coefficients of the metasurface, respectively. Note that $1+\tilde{r}=\tilde{t}$ holds owing to the continuity of the tangential components of the electric fields at $z=0$.
For a CPA, it is necessary that the outgoing waves destructively interfere in both $z<0$ and $z>0$, i.e., $\tilde{r}\tilde{E}_1+\tilde{t}\tilde{E}_2=\tilde{t}\tilde{E}_1+\tilde{r}\tilde{E}_2=0$ must be satisfied.
From this and $1+\tilde{r}=\tilde{t}$, we can derive the necessary conditions to realize a CPA as follows:
\begin{equation}
 \tilde{E}_1=\tilde{E}_2,
\end{equation}
\begin{equation}
 \tilde{t}=-\tilde{r}=\frac{1}{2}. \label{eq:Condition_meta}
\end{equation}
That is, the two incident waves must be in phase at the position of the metasurface and have equal intensity, and the metasurface must show 50\% absorption for a single incident wave, which is the maximum absorption possible for a thin metallic sheet~\cite{Woltersdorff1934}.
Assuming $\tilde{E}_2/\tilde{E}_1 = \gamma\exp(i \phi)$ with $\gamma\geq 0$, the absorption of the metasurface normalized to the total input power can be written as
\begin{equation}
 A=\frac{1}{2}+\frac{\gamma}{1+\gamma^2}\cos \phi. \label{eq:AmpPhi-Abs}
\end{equation}
In particular for $|\tilde{E}_1|=|\tilde{E}_2|$ or $\gamma=1$, we have
\begin{equation}
 A=\frac{1}{2}(1+\cos \phi). \label{eq:Phi-Abs}
\end{equation}
The relative phases $\phi=0$ and $\phi=\pi$ correspond to coherent perfect absorption and coherent perfect transmission, respectively. For $\phi=\pi/2$, the interference term in Eq.~(\ref{eq:Phi-Abs}) vanishes as in the case where $\phi$ is random or the two waves are incoherent.

\begin{figure}[t]
\centering
\includegraphics[width=8.0cm]{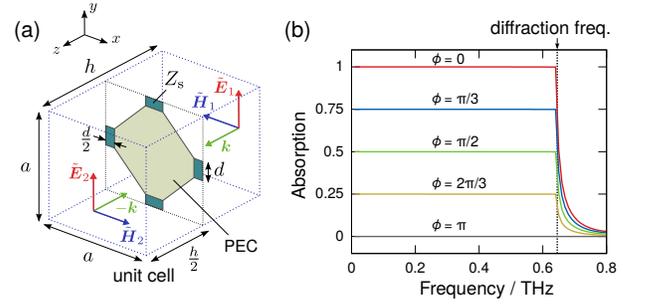}
\caption{(a)~Schematic of the simulation setup. The design parameters are: $a=150\,\upmu\mathrm{m}$, $d=30\,\upmu \mathrm{m}$, and $h=300\,\upmu\mathrm{m}$. (b)~Simulated absorption spectra for several relative phase differences $\phi$ of the two incident waves. The absorption is normalized to the total input power.}
\label{fig:simu}
\end{figure}

According to Babinet's principle, $\tilde{t}+\tilde{t}^{\mathrm{(comp)}}=1$ holds, where $\tilde{t}^{\mathrm{(comp)}}$ denotes the complex transmission coefficient for the complementary metasurface. Therefore, if the metasurface is composed of a self-complementary structure with $m$-fold symmetry ($m\geq 3$), then $\tilde{t}=\tilde{t}^{\mathrm{(comp)}}=1/2$~\cite{Nakata2013}. In other words, Eq.~(\ref{eq:Condition_meta}) is satisfied  for such metasurfaces, independent of frequency.
A trivial example is an unstructured absorptive thin film with a sheet impedance of $Z_\mathrm{s}=Z_0/(2n)$, which is equivalent to the thin-film CPA discussed in Ref.~\cite{Pu2012}.
Similarly, the resistive checkerboard-like structure with $4$-fold symmetry depicted in Fig.~\ref{fig:theory}(b) also satisfies Eq.~(\ref{eq:Condition_meta}), independent of frequency, due to its self-complementarity~\cite{Nakata2013,Urade2015}.
Above its lowest diffraction frequency $f_0$, however, perfect absorption does not occur because scattering into higher-order diffraction modes is allowed.
Note that absorption occurs only in the resistive sheets of the metasurface, which are the sole dissipative elements in this system, and that, in principle, the size of the resistive sheets can be  made arbitrarily small as long as self-complementarity holds.

To validate the above theoretical considerations, we performed numerical simulations with a commercial finite-element method solver~(\textsc{Ansys} \textsc{Hfss}).
Figure~\ref{fig:simu}(a) shows the setup of the simulation.
The design parameters were: $a=150\,\upmu\mathrm{m}$, $d=30\,\upmu \mathrm{m}$, and $h=300\,\upmu\mathrm{m}$.
The unit cell of the self-complementary checkerboard-like metasurface was placed at $z=0$, and the remaining part of the simulation domain was made of a dielectric medium with refractive index of $n=3.1$, which corresponds to the refractive index of sapphire for an ordinary ray at terahertz frequencies~\cite{Grischkowsky1990}.
The metasurface was composed of infinitely thin perfect electric conductor~(PEC) sheets and lossy resistive sheets with a sheet impedance of $Z_\mathrm{s}=Z_0/(2n)$.
The metasurface was excited by two plane waves incoming from excitation Floquet ports placed at the back~($z=-h/2$) and front~($z=h/2$) boundaries with complex electric-field amplitudes of $\tilde{E}_1$ and $\tilde{E}_2=\tilde{E}_1\exp(i \phi)$, respectively.
We simulated an infinitely periodic system by imposing periodic boundary conditions on the side boundaries of the simulation domain.
Higher-order diffraction modes were also taken into account in the simulation.
The absorption by the metasurface was calculated by integrating the surface loss density over all the resistive sheets in the unit cell.

The calculated absorption spectra normalized to the total input power are shown in Fig.~\ref{fig:simu}(b) for several values of $\phi$.
It is confirmed that perfect absorption occurs when $\phi=0$.
As $\phi$ increases, the absorption decreases and vanishes at $\phi=\pi$.
Note that the absorption spectrum for $\phi=\pi/2$ is the same as in the case of single-wave illumination.
The dependence of the absorption on $\phi$ agrees well with Eq.~(\ref{eq:Phi-Abs}).
Moreover, the absorption is nearly independent of frequency below the lowest diffraction frequency, $f_0=0.65\,\mathrm{THz}$, of the metasurface, while it becomes small above $f_0$.
These tendencies are  in agreement with the above theoretical analysis.
The numerical results verify that the self-complementary checkerboard-like metasurface behaves like a CPA when symmetrically illuminated by two waves.

\begin{figure}[bt]
\centering
\includegraphics[width=8cm]{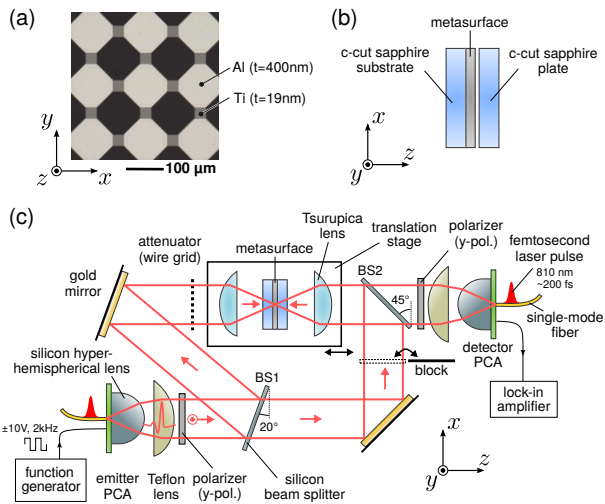}
\caption{(a)~Photomicrograph of the fabricated checkerboard-like metasurface. (b)~Setup of the metasurface during the measurements. (c)~Schematic of the experimental setup. The optical part is not depicted.}
\label{fig:exp}
\end{figure}

Finally, we experimentally demonstrate the CPA based on a self-complementary metasurface in the terahertz regime.
A self-complementary checkerboard-like metasurface was fabricated on a c-cut sapphire substrate~($20\times 20\times 0.9\,\mathrm{mm}^3$) using the direct-write laser lithography~(DWL2000, Heiderberg Instruments Mikrotechnik), electron beam evaporation, and lift-off process for titanium~(thickness: $19\,\mathrm{nm}$) and aluminum~(thickness: $400\,\mathrm{nm}$) layers repeatedly. The detailed fabrication process is described in Ref.~\cite{Urade2015}. Figure~\ref{fig:exp}(a) shows a photomicrograph of the prepared sample. The design parameters were the same as those in Fig.~\ref{fig:simu}(a). The thin titanium sheets serve as the resistive sheets with a sheet impedance of $Z_\mathrm{s}=Z_0/(2n)$~\cite{Urade2015}, where the sheet impedance depends on the thickness of the sheet.
Another c-cut sapphire plate with the same thickness as the substrate was placed on the metasurface so that the metasurface was sandwiched between the two sapphire plates, as depicted in Fig.~\ref{fig:exp}(b). This assures approximate mirror symmetry with respect to the metasurface, which is essential for Babinet's principle and therefore satisfying Eq.~(\ref{eq:Condition_meta})~\cite{Urade2015}.
Note that the anisotropy of the dielectric property also invalidates Babinet's principle because electromagnetic duality is broken. In the presented case, the c-cut sapphire substrate has a higher refractive index by $0.3$ in the $z$ direction~\cite{Grischkowsky1990}.
However, this effect is negligible except in the vicinity of the diffraction frequency.
In fact, it has been confirmed that the transmission spectrum of the fabricated sample satisfies $\tilde{t}(\omega)\approx 1/2$ in Ref.~\cite{Urade2015}.

The experimental setup is depicted in Fig.~\ref{fig:exp}(c). The system is based on a conventional terahertz time-domain spectroscopy system~\cite{Grischkowsky1990} with fiber-coupled photoconductive antennas~(PCAs) to emit and detect terahertz pulses. 
The group delay dispersion of the fibers, which causes significant pulse broadening, was compensated for by inserting grating pairs before laser-to-fiber coupling~\cite{Crooker2002}.
The emitter antenna generated terahertz pulses which were linearly polarized in the $y$ direction~[Fig.~\ref{fig:exp}(c)]. The emitted terahertz waves were collimated into a $22$-$\textrm{mm}$ diameter beam via a silicon hyper-hemispherical lens and a Teflon lens and then divided into two paths by a beam splitter~(BS1) made of high-resistivity silicon~(thickness: $2.5\,\mathrm{mm}$).
The pulses reflected at BS1 were focused onto the metasurface in the positive $z$ direction using a Tsurupica lens~(focal length: $50\,\textrm{mm}$). A wire grid polarizer was inserted before the Tsurupica lens to attenuate the $y$ component of the transmitted pulses, so that the pulses in the two paths had equal intensities at the metasurface.
The amount of attenuation was determined based on the Fresnel equations of the air-silicon interface of the beam splitters assuming that the silicon's refractive index is 3.4~\cite{Grischkowsky1990}.
The pulses transmitted through BS1 were reflected by a gold mirror and another silicon beam splitter~(BS2) to assure in-phase illumination onto the metasurface and then focused onto the metasurface in the negative $z$ direction.
The metasurface and the Tsurupica lenses were mounted on a motorized linear translation stage, which enables adjustments of the two terahertz path lengths.
The detector antenna measured the $y$ component of the electric field of the terahertz pulses, which were transmitted through or reflected at the metasurface, using the lock-in detection technique.
The temporal waveforms of the terahertz electric field were obtained as the average of 150 scans.
All measurements were performed under dry-air-purged conditions.
For the post processing, a $20$-$\mathrm{ps}$ width time window was used to separate the signals of interest from the preceding wave caused by the reflection at the air-sapphire boundary, and the succeeding echo pulses caused by multiple reflections inside the silicon beam splitters and sapphire plates.
Finally, note that it is sufficient to detect terahertz signals on one side of the checkerboard metasurface in order to demonstrate its CPA behavior. 
This is because the power of the terahertz signal is equal on both sides of the metasurface due to the symmetrical configuration of the illumination and the metasurface.
We also note that we obtained almost the same results as the following data when the sample was reversed.

\begin{figure}[t]
\centering
\includegraphics[width=8.0cm]{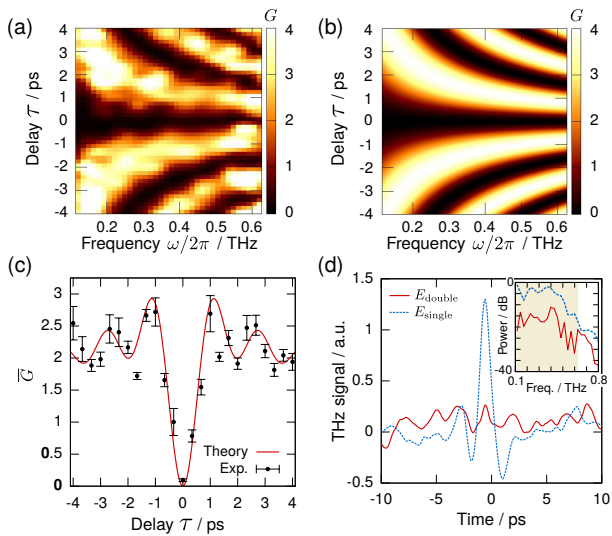}
\caption{(a)~Experimentally obtained $G(\omega, \tau)$. (b)~Theoretical $G(\omega, \tau)$. (c)~Comparison of the theoretical and experimental $\overline{G}(\tau)$. The error bars on the experimental values indicate two standard errors. (d)~$E_\mathrm{single}(t)$ and $E_\mathrm{double}(t)$ for $\tau =0$. The inset shows the power spectra of them.}
\label{fig:result}
\end{figure}

While changing the position of the translation stage with a step of $50\,\upmu\mathrm{m}$, we measured the terahertz electric fields for (i) single-sided pulse illumination, $E_\mathrm{single}(t)$, where one of the paths was blocked by a metallic sheet, as shown in Fig.~\ref{fig:exp}(c), and (ii) double-sided pulse illumination, $E_\mathrm{double}(t)$.
To evaluate the effect of the double-sided illumination, which induces coherent perfect absorption, we calculated $G(\omega,\tau):=|\tilde{E}_\mathrm{double}(\omega)/\tilde{E}_\mathrm{single}(\omega)|^2$, where $\tau$ denotes the delay time between the two pulses at the metasurface and $\tilde{E}$ represents the Fourier-transforms of the time-domain data.
A decrease in $G(\omega,\tau)$ indicates destructive interference of the outgoing waves from the metasurface caused by the double-sided excitation.
Figure~\ref{fig:result}(a) shows the experimentally obtained $G(\omega,\tau)$. We define $\tau=0$ as the delay position in which the average over frequency $\overline{G}(\tau)$ has a minimum. Here, $\overline{G}(\tau):=\int_{\omega_1}^{\omega_2} G(\omega,\tau)\mathrm{d}\omega / (\omega_2-\omega_1)$ with $\omega_1/(2\pi)=0.11\,\mathrm{THz}$ and $\omega_2/(2\pi)=0.63\,\mathrm{THz}$.
We observe a fringe pattern, which indicates that constructive and destructive interference alternately occur depending on the incident frequency and the delay time.
The broadband depression of $G(\omega,\tau)$ near $\tau=0$ suggests the CPA behavior of this system.
In theory, by assuming $\tilde{t}=-\tilde{r}=1/2$ and $\tilde{E}_2(\omega)=\exp(i \omega \tau)\tilde{E}_1(\omega)$, we can analytically calculate $G(\omega,\tau)$:
\begin{equation}
 G(\omega,\tau)= 2[1-\cos(\omega\tau)].
\end{equation}
Figure~\ref{fig:result}(b) depicts the theoretical $G(\omega,\tau)$.
It is confirmed that the fringe pattern qualitatively agrees with that of the experiment.
Moreover, Fig.~\ref{fig:result}(c) shows a comparison of the $\overline{G}(\tau)$ values found via the theory and the experiment, with error bars showing standard errors that propagate from the time domain data~\cite{Withayachumnankul2008a}. The experimental result shows quantitative agreement with the theoretical curve.
Figure~\ref{fig:result}(d) shows $E_\mathrm{single}(t)$ and $E_\mathrm{double}(t)$ for $\tau =0$. The inset shows the power spectra  of them. We can confirm a decrease in the terahertz pulse amplitude  for the double-sided illumination, which clearly indicates the broadband nature of the proposed CPA in the time domain.
Therefore, we conclude that the CPA behavior of the self-complementary checkerboard-like metasurface was experimentally demonstrated.
From $\overline{G}(\tau)$, the experimental maximum absorption is estimated to be approximately $98\%$ if we assume $\tilde{t}=1/2$.

In conclusion, we  demonstrated that a self-complementary checkerboard-like metasurface works as a coherent perfect absorber. In addition to the theoretical demonstration, numerical simulations and  experiments at terahertz frequencies were presented.
The proposed CPA can operate in a broad bandwidth below its lowest diffraction frequency.
In addition, regions that contribute to the absorption are much smaller than the wavelength of the incident electromagnetic waves.
For the presented design, the absorptive area amounts to only $8\%$ of the unit cell and can be further decreased in principle. 
The compatibility of the deep-subwavelength energy concentration with the broadband operation  is applicable to ultrafast Joule heating of subwavelength regions by intense terahertz transients, which can enhance nonlinear effects caused by temperature increases in the resistive sheets.
The temperature increase in the resistive sheets is also applicable to thermal imaging of terahertz beam profile~\cite{Tadokoro2015} in a phase-sensitive manner.
Further deep-subwavelength absorption is possible using graphene for the resistive parts of the metasurface~\cite{Fan2014}.

\bigskip
\noindent{\textbf{Funding.}}\indent
JSPS KAKENHI Grant Numbers JP15J07603, JP22109004.

\bigskip
\noindent{\textbf{Acknowledgment.}}\indent
The sample was prepared using the instruments at the Kyoto University Nano Technology Hub, as part of the ``Nanotechnology Platform Project'' sponsored by MEXT in Japan.
One of the authors~(Y.~Urade) was supported by a JSPS Research Fellowship for Young Scientists.


\end{document}